\documentclass[twocolumn]{emulateapj}

\usepackage{natbib} 
\usepackage{epsfig} 
\usepackage{lscape} 
\usepackage{graphicx} 
\slugcomment{Revised Version July 1, 2009}

\shorttitle{Observational Window Functions in Planet Transit Surveys}

\shortauthors{von Braun et al.}

\begin{document}

\title{Observational Window Functions in Planet Transit Surveys}

\author{Kaspar von Braun}
\author{Stephen R. Kane}
\author{David R. Ciardi}

\affil{NASA Exoplanet Science Institute, 
California Institute of Technology, 
MC 100-22, Pasadena, CA 91125}
\email{kaspar, skane, ciardi@ipac.caltech.edu}


\begin{abstract}

The probability that an existing planetary transit is detectable in one's data is sensitively dependent upon the window function of the observations. We quantitatively characterize and provide visualizations of the dependence of this probability as a function of orbital period upon several observing strategy and astrophysical parameters, such as length of observing run, observing cadence, length of night, transit duration and depth, and the minimum number of sampled transits.  The ability to detect a transit is directly related to the intrinsic noise of the observations. In our simulations of observational window functions, we explicitly address non-correlated (gaussian or white) noise and correlated (red) noise and discuss how these two noise components affect transit detectability in fundamentally different manners, especially for long periods and/or small transit depths. We furthermore discuss the consequence of competing effects on transit detectability, elaborate on measures of observing strategies, and examine the projected efficiency of different transit survey scenarios with respect to certain regions of parameter space. 

\end{abstract}


\keywords{eclipses --- methods: statistical ---  planetary systems --- surveys
--- techniques: photometric --- time}


\section{Introduction}\label{introduction}

The signal-to-noise ratio (SNR) of a planetary transit detection in
photometric time series data can, in the simplest case, be approximated by:

\begin{equation}
SNR_{transit} = \frac{depth}{\sigma}\sqrt{n}.
\label{simple_equation}
\end{equation}

\noindent In this equation, $depth$ is the transit depth in magnitudes, $\sigma$ 
represents the photometric measurement uncertainty in magnitudes per data
point (assumed here to be the same for all data points), and $n$ equals the total
number of data points observed during transit \citep{p06}. One essential
assumption in this equation is the absence of any statistically correlated
(red) noise, i.e., only random (white) noise is present.  White noise is
defined as noise that is uncorrelated from data point to data point; typical
sources are photon noise and sky background noise. The relative contribution
of white noise to the total noise decreases with increasing brightness of the
observed target and number of data points.

\citet{p06} and \citet{pzq06} showed that calculations of transit
SNRs with only white noise, as in Equation \ref{simple_equation}, is
often insufficient and overly optimistic. Instead, one needs to account for the presence of red noise in
calculations of SNRs and the corresponding yield projections for transit
surveys.  Red noise is defined as noise that is correlated from data point to
data point; it is not necessarily removed through standard differential or
ensemble photometry techniques. Typical sources of red noise may be weather,
seeing changes, tracking/guiding errors, flatfielding errors,
changes in airmass, or intrinsic astrophysical changes in target
brightness. It does not change as a function of target magnitude, and is
generally independent of the number of observational data points \citep[see
equation 9 in][]{pzq06}. Thus, planetary transit searches are particularly
sensitive to red noise, due to their focus on bright targets and high number of
observational epochs: both are aimed to reduce white noise, and therefore make red noise the dominant component.

A detailed description of the transit detection SNR which includes both
white and red noise components is given by \citet{pzq06}:

\begin{eqnarray}
SNR_{transit} &=& 
\frac{ depth }{\sqrt{\frac{1}{n^2} \sum_{i,j}cov[i;j] }} \nonumber \\
&=&
\frac{ depth }{\sqrt{\frac{\sigma^2}{n} + 
\frac{1}{n^2}\sum_{i\neq j}cov[i;j] }},
\label{cov_equation}
\end{eqnarray}
 
\noindent where $cov[i;j]$ is the covariance matrix with elements $C_{ij}$
representing the correlation coefficents between the $i$-th and $j$-th
measurements obtained during transit. All diagonal elements $C_{ii}=\sigma^2_i$
are not correlated with other measurements and thus represent the uncorrelated
or white noise uncertainties in the $i$-th measurement. These diagonal
elements are assumed to be the same, i.e., $\sigma_i = \sigma$ for all values
of $i$.

In order to make the above equation more practically calculable, \citet{pzq06}
assume that statistical correlation among data points from different transits
will be much weaker than among data points observed during the same
transit. They furthermore separate the total noise into a purely uncorrelated
(white) component $\sigma_w$ and a purely correlated (red) component
$\sigma_r$ and derive an approximation of Equation \ref{cov_equation}:

\begin{equation}
SNR_{transit} = \sqrt{
\frac{ \left ( depth \cdot n \right )^{2}}{\sum^{N_{tr}}_{k=1}\left[ n^2_k \left ( \frac{\sigma^2_w}{n_k} +
\sigma^2_r \right ) \right ] } },
\label{approximation_equation}
\end{equation}

\noindent where $n$ is the total number of data points observed during all
transits, $N_{tr}$ is the total number of transits observed, $n_{k}$ is the
number of data points observed during the $k$-th transit, and $\sigma_w$ and
$\sigma_r$ are the white and red noise components, respectively.

By means of Equations \ref{simple_equation} and \ref{approximation_equation},
it is clear that a planet transit SNR can be regarded as a function of transit
survey strategy and astrophysical parameters (see \S \ref{algorithm}). If this
SNR exceeds a certain threshold value, then an existing transiting planet is,
for the purposes of this paper, defined to be detectable in the data\footnote{We note that, to maximize applicability for astronomical planet transit surveys, we follow the arguments outlined above and in \citet{pzq06}, rather than using more rigorous treatments employed in the large body of statistical literature devoted to time-series analysis. These treatments include autoregressive moving-average (ARMA) modeling where events such as eclipses in the presence of red noise are sought with the help of autocorrelation functions at different time lags \citep{kl03a,kl03b,r05}, power spectrum analysis \citep{kt07}, or the use of surrogate data sets \citep{t98}.}.

The window function determines the probability, as a function of planetary
orbital period, that SNR$_{transit}$ exceeds this threshold.  In this paper, we examine the dependence of the
detection probability upon several astrophysical and transit survey strategy
parameters for a number of white noise and red noise assumptions as well as
criteria based on a minimum number of transits sampled.

Since our calculations are based on {\it existing} transits, we note that the
following aspects are not taken into account: the estimated frequency of
transiting exoplanets, any non-circular orbits, multi planet systems, and
detection of secondary transits. We also do not address the problem of false
positives and how to weed them out.  For more detailed studies of the above,
we refer the reader to the following studies: frequency of (transiting)
exoplanets: \citet{gdg06} and \citet{cbm08}; transit probability as a function
of orbital elements: \citet{barnes07}, \citet{burke08}, and \citet{kb08}; plus
see \citet{g07} and \citet{bg08} and references therein for a comprehensive
study of all factors influencing planet detections in transit
surveys.



We briefly outline our methods in \S \ref{algorithm}, which describes our algorithm in \S \ref{description_algorithm}, along with a
justification for the threshold SNR selection, and addresses
the respective influences of varying white and red noise components (\S
\ref{noises}), as well as the consideration of sampling at least $N_{tr}$
transits with one's data to constitute a detection (\S \ref{transit_number}).
We examine the effects of various survey strategy and
astrophysical parameters in \S \ref{parameters}.
Section \ref{surveys} contains the application of window functions for
selected scenarios and types of survey. We summarize and conclude in \S
\ref{conclusion}. 


\section{Algorithm and Parameters}
\label{algorithm}

In this Section, we provide a brief description of our algorithm and explain
our choices of the globally used values of input parameters in our
calculations in \S \ref{parameters}.


\subsection{Description of the Algorithm}
\label{description_algorithm}



The window function algorithm used in this paper is based on counting data
points observed during transit whose contribution to a virtual detection is
dependent on the values of $\sigma_w$ and $\sigma_r$ as defined in Equations \ref{cov_equation} and \ref{approximation_equation}, 
typically
measurable or calculable quantities in photometric time series surveys.

User-provided observing cadence, number of nights, and typical length of night are used to
generate an observing time line. From the input stellar and planetary
radii, we calculate transit depth and duration according to the equations in
\citet{sm03} (except in \S \ref{transit_duration} where we explicitly set
transit depth and duration), thereby assuming a central transit (i.e., impact parameter $b=0$) and
zero-length ingress and egress.  For each orbital period, a family of light
curves is generated for a range of starting phase angles; each with transits of user-defined photometric depths at the appropriate intervals.  

In the simulations, the number of data points per transit ($n_k$), number of
transits ($N_{tr}$) and total number of data points within all transits ($n$)
are tracked. It should be noted that an observation has to fit fully within a transit to be counted toward $n$ and $n_k$ (that is, it needs to start after the beginning of the transit and terminate before the end of the transit), resulting in shorter exposure times' being more favorable for transit detection in this algorithm. For every light curve, the SNR (Equations \ref{simple_equation}
and \ref{approximation_equation}) is calculated. If, for a given phase angle,
the SNR exceeds SNR$_{threshold}$, a transit is considered ``detected''. The
probability of detection (P$_{detection}$) for a given orbital period is
simply the ratio of phase angles for which a transit was detected to the total
number of phase angles.

Typical observational parameter values assumed in this paper (unless specifically noted) are: 
minutes for the observing cadence, one minute for the exposure time, tens of nights for observing run length,
and few to ten hours for the typical time of observation spent during one
night on the monitored target. Astrophysical parameter values are assumed to
be around 1.0 and 0.1 solar radii for the parent star and orbiting planet,
respectively, resulting in a transit depth of 0.01 mag. Transit duration
depends on period, but typical duty cycles are in the 1\% to few \% range.
Additionally, we set  $\sigma_w$ and $\sigma_r$ to a few millimagnitudes (mmag). The threshold
SNR is set to 7.0, based on the arguments in \citet{jcb02} and specifically \citet{pzq06,pab07}, which each use thresholds of 7--9 as acceptable values for reducing false-alarms
whilst maximizing real detections given a typical transit survey
configuration.

We note that, in contrast to other some window function calculations in the astronomical literature, we only
use the SNR criterion to quantify detections, along with an assumed minimum
number of sampled transits, and we do not require that, e.g., a full transit
be contained in the data \citep[as in, e.g., ][]{msy03,bls05}. We do not account for holes in the observing due to weather, telescope outages, or technical problems. Furthermore,
as mentioned in \S \ref{introduction}, we only calculate the probability of
detecting existiting primary transits in circular orbits. Finally, we assume
that the number of out-of-transit data points sampled is much higher than the
number of in-transit data points.

\subsection{Red Noise and White Noise}
\label{noises}

As outlined in \citet{pg05,pzq06,ap07,bc08}, red noise is the dominant source
of noise in the regimes of brightness and number of observational epochs where
transit surveys typically operate since red noise is independent of target
brightness. We show this effect in \S \ref{red_white_noise}. See \citet{ap07}
and \citet{iia07} for an in-depth discussion of different noise properties and
their calculations.

Typical ground-based survey estimates of $\sigma_r$, as defined in \S
\ref{introduction}, are on the order of 2--6 millimagnitudes (mmag)
\citep[e.g., ][]{pzq06,iia07,kcw07, nc08}. When subjected to detrending
algorithms such as TFA \citep{kgn05,kb07} or SYSREM \citep{tmz07}, $\sigma_r$
can be reduced to 1--2 mmag.

It is worth pointing out that the influence of red noise is much less of a
problem for targeted observations such as characterization of known planetary
transits \citep[see][for example]{gad09}. 

Studies to date \citep[e.g.,][]{aco08} have shown that the red noise in the
space-based CoRoT mission \citep{bab06} is significantly lower than in ground-based
counterparts, due in large part to the ``removal of the atmosphere''
\citep{pzq06,bg08}. Thus, the cause of any space based red noise not due to stellar
variations is most likely caused by variations in the thermal environment of
the spacecraft and detectors. Typical values for $\sigma_r$ in CoRoT light
curves are on the order of 0.5 mmag \citep[R. Alonso 2008, private communication; see also][]{apf09}.


\subsection{Number of Sampled Transits}
\label{transit_number}

One criterion often used to calculate detection efficiency and related survey
yield is the minimum number of sampled transits (i.e., the minimum number of
transits during which any data were obtained). 
An important factor in the success of the widely used BLS algorithm \citep{kzm02} is the initial folding of the data by a test period and
subsequent search for transit-like features in the phased data. Thus, its
power is really only realized for data that contain more than one sampled
transit.  We assume in this publication that the BLS algorithm has become an
``industry standard'' in the search for planetary transits, and we thus
require the existence of at least two transits in the data for a transit detection, except for where we explicitly change this criterion (\S
\ref{min_transit_number}). It is worth noting that
different simulations in the literature require different minimum numbers of
transits sampled, such as three for \citet{pzq06}.


\section{The Influence of Observational Window Functions and Astrophysical Parameters on Transit Detection Probability}
\label{parameters}

Careful consideration of the various strategy aspects involved in planetary transit surveys and a number of astrophysical parameters will have significant effects on the detection efficiency of existing transits \citep{msy03,bls05,pg05,bg08,bc08,b09}. This Section quantitatively illustrates these effects under consideration of the assumptions described in \S \ref{algorithm}. For the sake of clarity, we vary one parameter at a time, leaving all others fixed to values justified in \S \ref{introduction} and \S \ref{algorithm}. In particular, \S \ref{red_white_noise} examines different values for red and white noise, \S \ref{run_length} looks at various observing run lengths (with a given number of hours of observing per night), whereas \S \ref{night_length} assumes a number of consecutive observing nights but varies their lengths. \S \ref{cadence} investigates different observing cadences. In \S \ref{min_transit_number}, we explicitly change the criterion of a minimum of two transits sampled for a detection that we mention in \S \ref{transit_number} to see how requiring a larger number decreases detection efficiency. \S \ref{transit_duration} deals with different transit depths and durations. 


The values of the parameters held constant in the respective calculation are given in the caption of the appropriate figure. The solid (blue) line shows the detection efficiency in the hypothetical case of zero red noise, and the dashed (red) line shows the same for a given $\sigma_r \neq 0$. The corresponding table shows the mean values of $P_{detection}$ for the ranges of orbital periods given in the first column under assumptions of the indicated magnitudes of  $\sigma_w$ and $\sigma_r$. 


\subsection{Amount of Red Noise and White Noise}
\label{red_white_noise}

\begin{figure*}
\begin{center}
 \includegraphics[angle=270,scale=0.6]{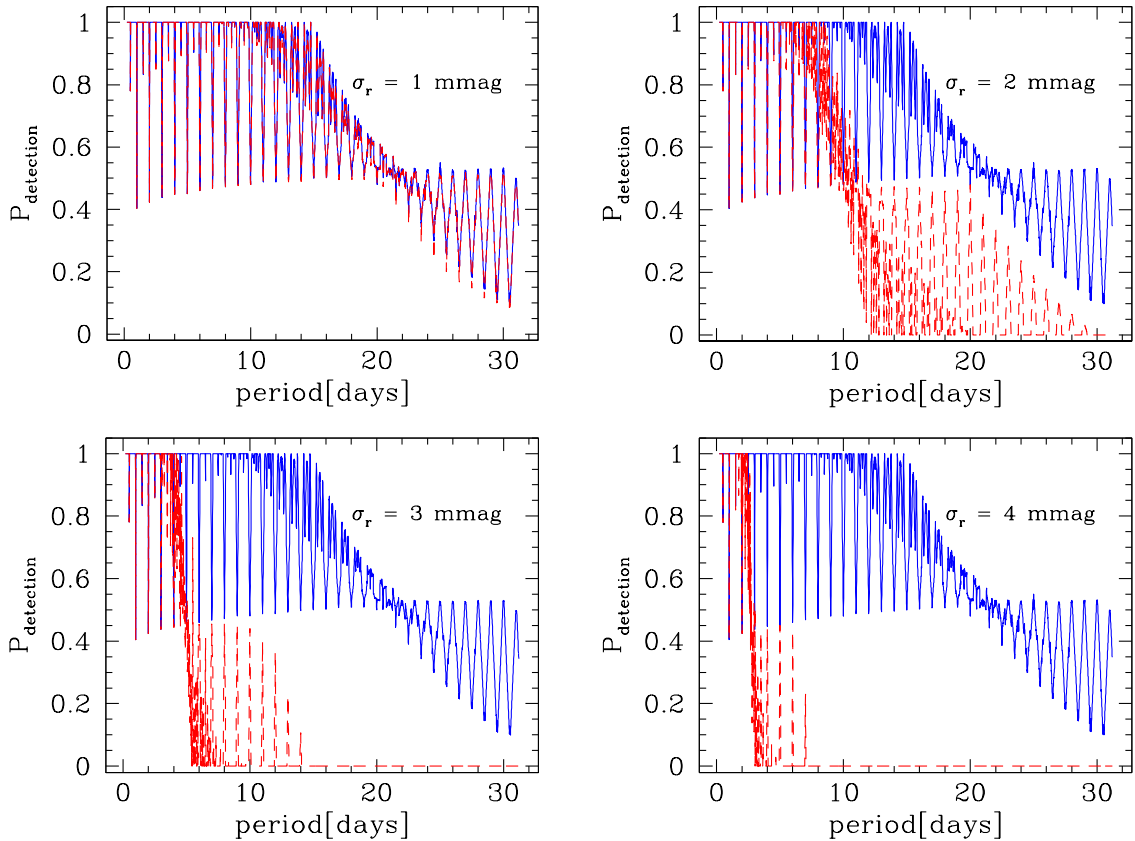} 
 \caption{Variable red noise and constant white noise ($\sigma_w = 5$
 mmag). The (blue) solid line indicates the detection efficiency for no red
 noise ($\sigma_r = 0$) and is shown in all four panels. The (red) dashed line
 indicates detection efficiency for varying levels of $\sigma_r$ as indicated
 in the respective panel. Additional parameters are: SNR$_{threshold}$ = 7.0,
 60 consecutive observing nights with 8 hours of observing every night, an
 observing cadence of 5 minutes, $R_{star} = R_{\odot}$, $R_{planet} = 0.1
 R_{\odot}$, $M_{star} = M_{\odot}$, $M_{planet} \ll M_{\odot}$. See Table \ref{tab_rednoise} for mean values of $P_{detection}$ over various period ranges, and \S \ref{red_white_noise} for details.}
 \label{fig_rednoise}
\end{center}
\end{figure*}

\begin{figure*}
\begin{center}
 \includegraphics[angle=270,scale=0.6]{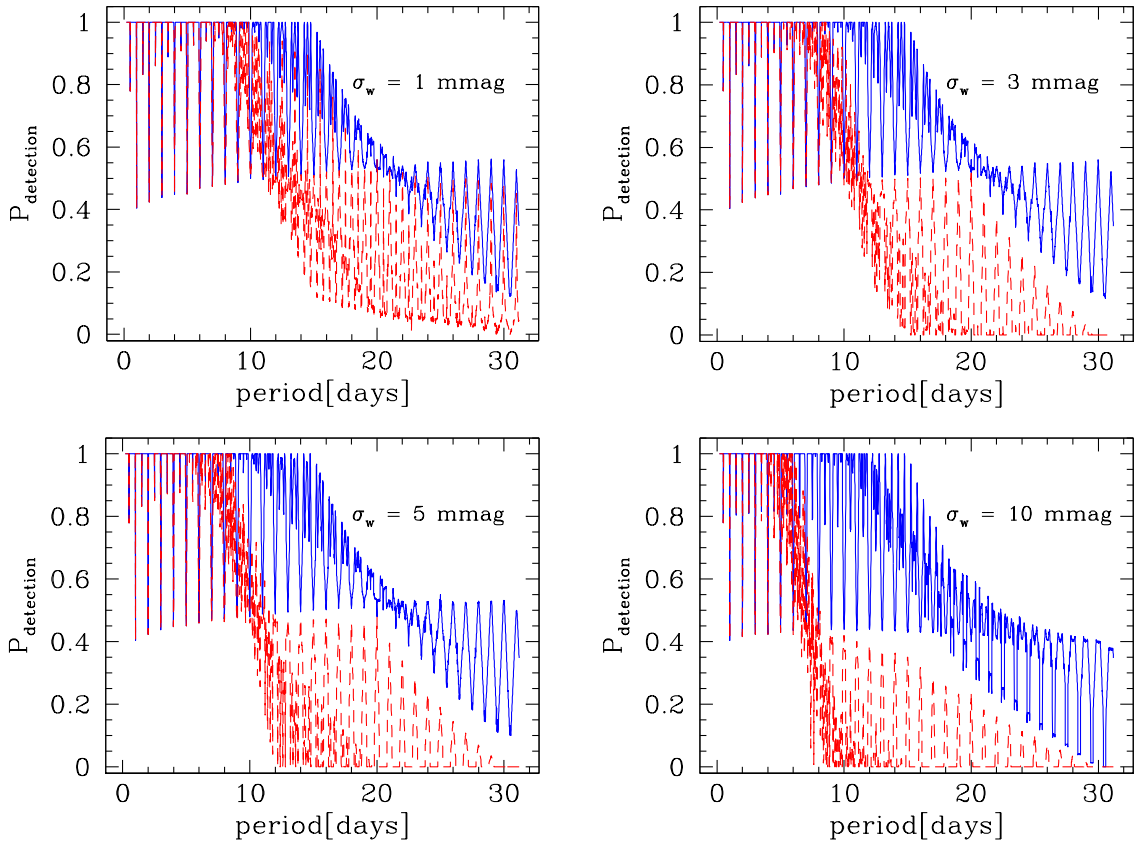} 
 \caption{Variable white noise and constant red noise. The value for
 $\sigma_w$ is given in the top right of every respective panel.  The (blue)
 solid line indicates the detection efficiency $\sigma_r = 0$, and the (red)
 dashed for $\sigma_r = 2$ mmag. Additional parameters are: SNR$_{threshold}$
 = 7.0, 60 consecutive observing nights with 8 hours of observing every night,
 an observing cadence of 5 minutes, $R_{star} = R_{\odot}$, $R_{planet} = 0.1
 R_{\odot}$, $M_{star} = M_{\odot}$, $M_{planet} \ll M_{\odot}$. See Table \ref{tab_whitenoise} for mean values of $P_{detection}$ over various period ranges, and \S \ref{red_white_noise} for discussion.}
 \label{fig_whitenoise}
\end{center}
\end{figure*}

The contribution of red noise is independent of target brightness (unlike white noise, which is mostly due to photon noise for the brightest targets). Since planet transits are typically detected around the brightest sources in a given data set, red noise will be the dominant source of noise \citep{pg05, pzq06, ap07, bg08}. Figures \ref{fig_rednoise} and \ref{fig_whitenoise} along with Tables \ref{tab_rednoise} and \ref{tab_whitenoise} quantitatively substantiate this statement, illustrating the influences of different amounts of red and white noises for different period ranges. As we mention in \S \ref{noises}, typical values for the wide-field ground-based transit surveys before detrending are $\sigma_w \sim$ 5 mmag and $\sigma_r \sim$ 2--6 mmag, which reduces to $\sigma_r \sim$ 1--2 mmag after detrending. 

The difference in $P_{detection}$ in Fig. \ref{fig_rednoise} and Table \ref{tab_rednoise} between $\sigma_r=1$ mmag and $\sigma_r=4$ mmag is very significant for longer periods. In addition, Fig. \ref{fig_whitenoise} and Table \ref{tab_whitenoise} show how small the influence of $\sigma_w$ upon $P_{detection}$ is for no or very little red noise. Thus, the value of minimizing the influences of red noise during observing (even at the expense of increasing $\sigma_w$ if necessary), and of applying detrending algorithms such as SYSREM \citep{tmz07} or TFA \citep{kgn05,kb07} to one's data as part of their reduction can hardly be overstated.

\begin{deluxetable}{lccccc}
\tablecaption{Mean $P_{detection}$ Values for Different Magnitudes of $\sigma_r$ \label{tab_rednoise}}
\tablewidth{0pc}
\tablehead{
\colhead{} &
\colhead{$\sigma_r = 0$} &
\colhead{1 mmag} &
\colhead{2 mmag} &
\colhead{3 mmag} &
\colhead{4 mmag} 
}
\startdata
0--5 days & 0.983 & 0.982 & 0.978 & 0.906 & 0.540\\
5--10 days & 0.945 & 0.942 & 0.819 & 0.102 & 0.010\\
10--20 days & 0.753 & 0.747 & 0.235 & 0.012 & 0.000\\
20--30 days & 0.425 & 0.421 & 0.051 & 0.000 & 0.000\\
\enddata
\tablecomments{Mean values for $P_{detection}$ for various period ranges (column 1) as a function of different values of $\sigma_r$ (Fig. \ref{fig_rednoise}). Assumed parameters are given in the caption of Fig. \ref{fig_rednoise}. For discussion, see \S \ref{red_white_noise}.}
\end{deluxetable}

\begin{deluxetable}{lcccc}
\tablecaption{Mean $P_{detection}$ Values for Different Magnitudes of $\sigma_w$ \label{tab_whitenoise}}
\tablewidth{0pc}
\tablecolumns{6}
\tablehead{
\colhead{$\sigma_w$} &
\colhead{1 mmag} &
\colhead{3 mmag} &
\colhead{5 mmag} &
\colhead{10 mmag} }
\startdata
\cutinhead{($\sigma_r = 0$)}
0--5 days &  0.984 & 0.983 & 0.983 & 0.977\\
5--10 days &  0.947 & 0.947 & 0.945 & 0.916\\
10--20 days &  0.755 & 0.755 & 0.753 & 0.677\\
20--30 days &  0.428 & 0.428 & 0.425 & 0.355 \\
\cutinhead{($\sigma_r = 2$mmag)}
0--5 days &  0.981 & 0.980 & 0.978 & 0.967\\
5--10 days &  0.897 & 0.857 & 0.819 & 0.444\\
10--20 days &  0.412 & 0.281 & 0.235 & 0.072\\
20--30 days &  0.131 & 0.058 & 0.051 & 0.017 \\
\enddata
\tablecomments{Mean values for $P_{detection}$ for various period ranges (column 1) as function of $\sigma_w$ (Fig. \ref{fig_whitenoise}). Assumed parameters are given in the caption of Fig. \ref{fig_whitenoise}. For discussion, see \S \ref{red_white_noise}.}
\end{deluxetable}


\subsection{Observing Run Length}
\label{run_length}

\begin{figure*}
\begin{center}
 \includegraphics[angle=270,scale=0.6]{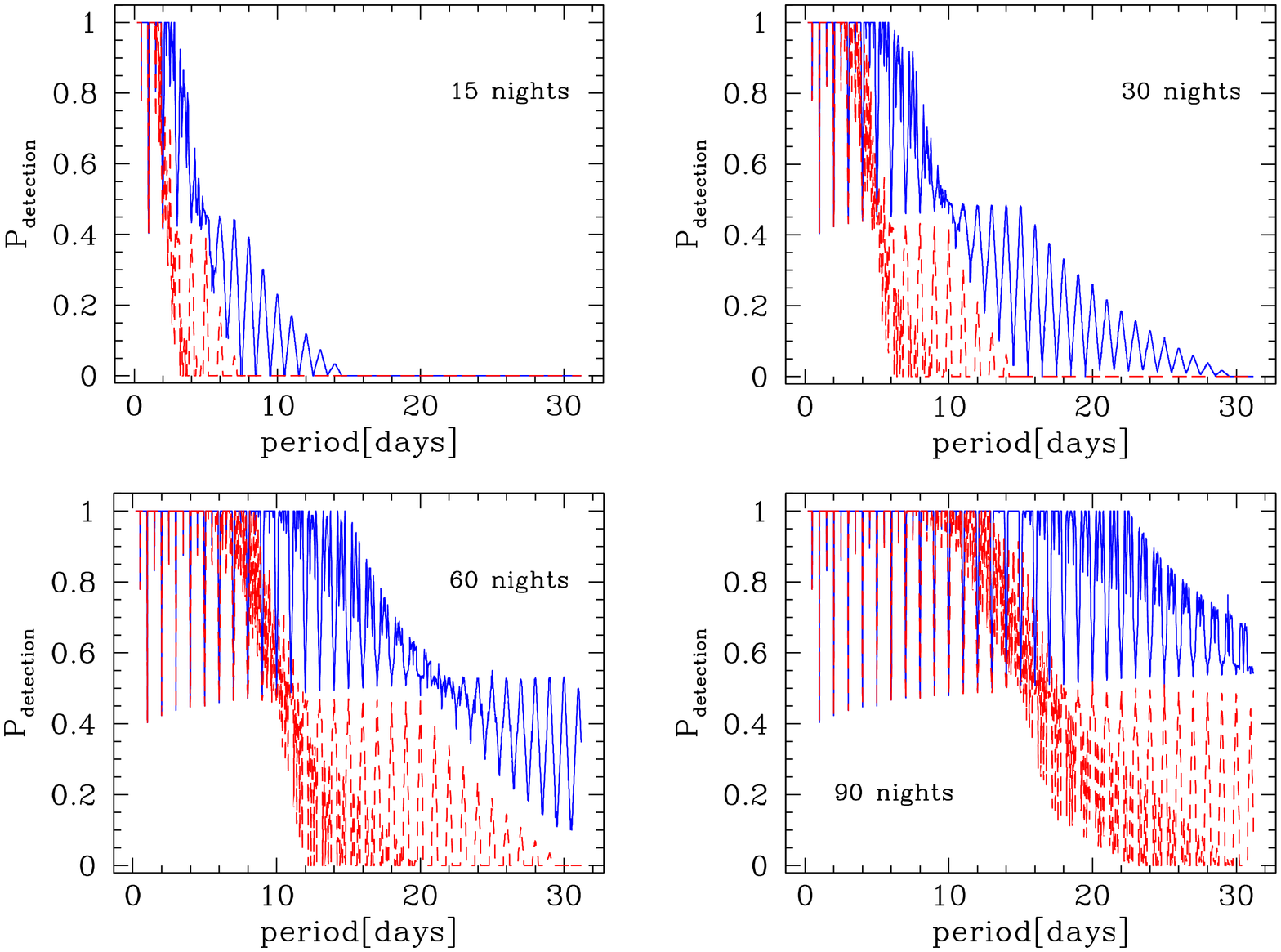} 
 \caption{The influence of length of observing run on the detection
 efficiency, shown for 15, 30, 60, and 90 nights (indicated in the respective
 panel).  $\sigma_w$ is assumed to be 5 mmag. The (blue) solid line indicates
 the detection efficiency $\sigma_r = 0$, and the (red) dashed for $\sigma_r =
 2$ mmag. Additional parameters are: SNR$_{threshold}$ = 7.0, 8 hours of
 observing every night, an observing cadence of 5 minutes, $R_{star} =
 R_{\odot}$, $R_{planet} = 0.1 R_{\odot}$, $M_{star} = M_{\odot}$, $M_{planet}
 \ll M_{\odot}$. See Table \ref{tab_obsrunlength} for mean values of $P_{detection}$ over various period ranges, and \S \ref{run_length} for details.}  \label{fig_obsrunlength}
\end{center}
\end{figure*}

For any kind of transit survey that has limited access to telescope time, the question of how long to spend on one field will occur at some point during the design of the observing strategy. At what point is it worth switching to a different field to increase the number of targets without overly reducing the probability of detecting existing transits in the data? 
We provide insight into the answer to this question in Fig. \ref{fig_obsrunlength} and Table \ref{tab_obsrunlength}. 

To first order, observing a field for few nights will yield an almost negligible probability of detection, potentially leading to a waste of telescope time. Alternatively, it may be wise not to stay on a single field for too long but rather to double the chances of detecting any planetary transits by switching fields and thus increasing the number of monitored stars. It is ultimately a question of the period range one is sampling in a given survey. As Fig. \ref{fig_obsrunlength} shows for a typical set of parameters, ``very hot Jupiters", i.e., planets with periods up to $\sim$3 days per the definition in \citet{gsm05}, can be detected even with a residual presence of red noise and ``only" 15 nights (eight hours per night) of monitoring. However, longer period planets ($\sim$6 days and longer) remain elusive (for $\sigma_r  \geq$ 2mmag) until the length of the observing run exceeds 30 nights. 

\begin{deluxetable}{lcccc}
\tablecaption{Mean $P_{detection}$ Values for Different Observing Run Lengths \label{tab_obsrunlength}}
\tablewidth{0pc}
\tablehead{
\colhead{Nights} &
\colhead{15} &
\colhead{30} &
\colhead{60} &
\colhead{90} }
\startdata
\cutinhead{($\sigma_r = 0$)}
0--5 days &  0.792 & 0.957 & 0.983 & 0.988\\
5--10 days &  0.238 & 0.696 & 0.945 & 0.969\\
10--20 days &  0.027 & 0.279 & 0.753 & 0.918\\
20--30 days &  0.000 & 0.064 & 0.425 & 0.736 \\
\cutinhead{($\sigma_r = 2$mmag)}
0--5 days &  0.482 & 0.865 & 0.978 & 0.987\\
5--10 days &  0.020 & 0.189 & 0.819 & 0.954\\
10--20 days &  0.000 & 0.022 & 0.235 & 0.618\\
20--30 days &  0.000 & 0.000 & 0.051 & 0.188 \\
\enddata
\tablecomments{Mean values for $P_{detection}$ for various period ranges (column 1) as function of observing run length (Fig. \ref{fig_obsrunlength}). Assumed parameters are given in the caption of Fig. \ref{fig_obsrunlength}. For discussion, see \S \ref{run_length}.}
\end{deluxetable}


\subsection{Length of Night}
\label{night_length}

The amount of time for which a given target field can be observed from the ground during one night depends on its celestial coordinates, the location of the telescope, the time of year, and, of course, outages due to weather, or technical or other problems. Special cases are discussed below, such as 
space-based observing (\S \ref{space}) or synoptic surveys (\S \ref{synoptic}). 

\begin{figure*}
\begin{center}
 \includegraphics[angle=270,scale=0.6]{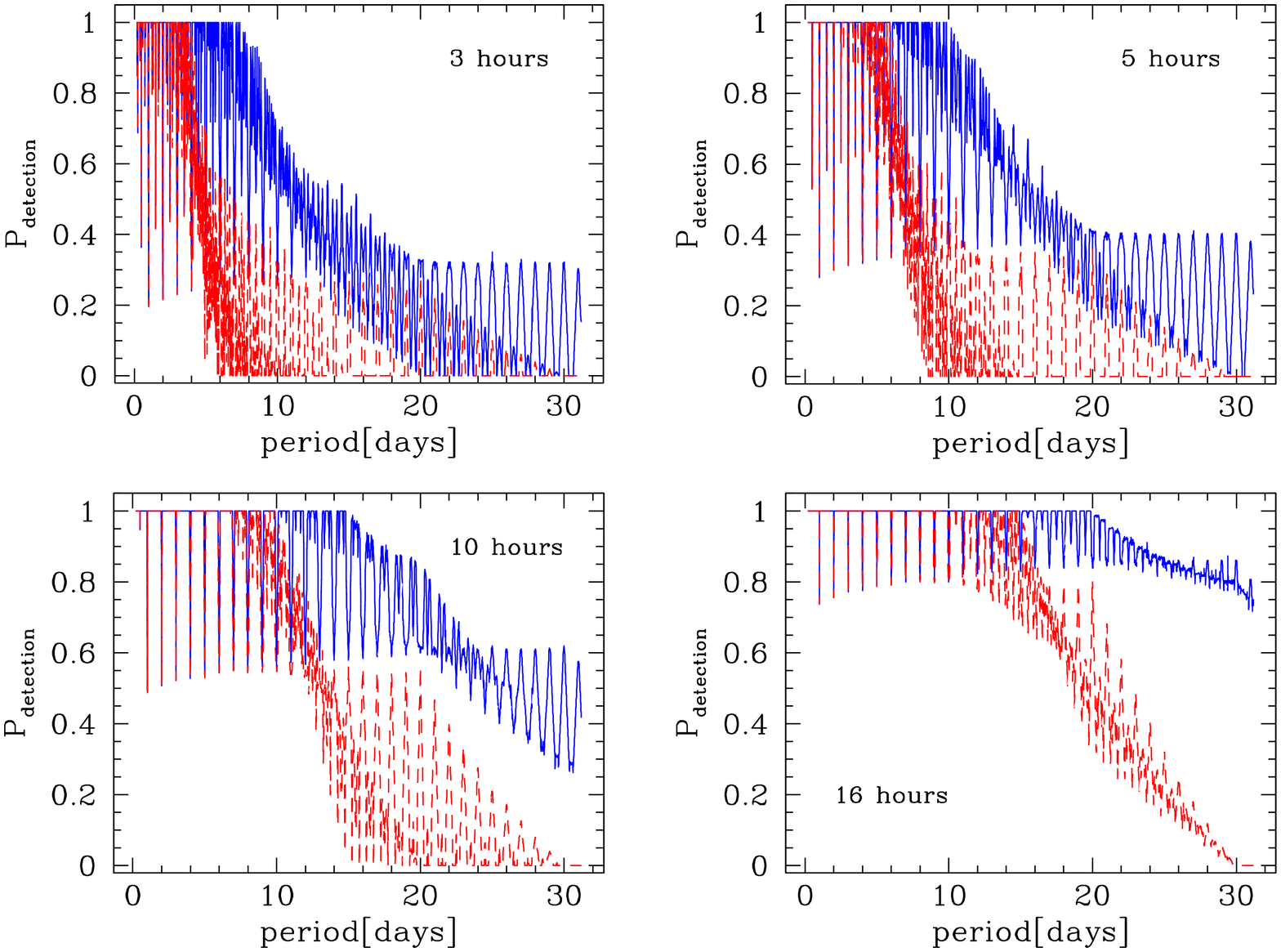} 
 \caption{The influence of the length of night on the detection efficiency,
 shown for 3, 5, 10, and 16 hours (indicated in the respective panel).
 $\sigma_w$ is assumed to be 5 mmag. The (blue) solid line indicates the
 detection efficiency $\sigma_r = 0$, and the (red) dashed for $\sigma_r = 2$
 mmag. Additional parameters are: SNR$_{threshold}$ = 7.0, a 5-minute
 observing cadence, 60 consecutive nights, $R_{star} = R_{\odot}$, $R_{planet}
 = 0.1 R_{\odot}$, $M_{star} = M_{\odot}$, $M_{planet} \ll M_{\odot}$. See Table \ref{tab_nightlength} for mean values of $P_{detection}$ over various period ranges, and \S \ref{night_length} for details.}
 \label{fig_nightlength}
\end{center}
\end{figure*}

The length of night can also depend on observing strategy. As an alternative to decreasing the number of nights spent on a single target field to increase the number of monitored stars (\S \ref{run_length}), one may instead choose to split the night up between two or more fields, thereby decreasing the number of hours spent on each one of them. We illustrate the effect of such strategies in Fig. \ref{fig_nightlength} and Table \ref{tab_nightlength} in which we assume basically the same parameters as for Figures \ref{fig_obsrunlength} and \ref{fig_cadence} (see \S \ref{cadence}) for purposes of comparison. The situation shown in the bottom right panel can obviously only be achieved at numerically high latitudes on Earth during the respective winter season, or from space, but serves as a comparison to the scenarios encountered in transit surveys conducted from moderate latitudes. As in \S \ref{run_length}, the choice of strategy depends on the range of periods that is probed. 

One expected and visible effect in Fig. \ref{fig_nightlength} is the decreasing depth of spikes in $P_{detection}$ with longer lengths of night as the diurnal cycle becomes less of a factor in transit detection. As evidenced in Figures \ref{fig_5nights} and \ref{fig_corot}, the spikes eventually disappear altogether when observing becomes uninterrupted, as, e.g., from space.

\begin{deluxetable}{lcccc}
\tablecaption{Mean $P_{detection}$ Values for Different Night Lengths \label{tab_nightlength}}
\tablewidth{0pc}
\tablehead{
\colhead{Hours/Night} &
\colhead{3} &
\colhead{5} &
\colhead{10} &
\colhead{16} }
\startdata
\cutinhead{($\sigma_r = 0$)}
0--5 days &  0.949 & 0.970 & 0.989 & 0.997\\
5--10 days &  0.764 & 0.886 & 0.962 & 0.992\\
10--20 days &  0.358 & 0.543 & 0.846 & 0.979\\
20--30 days &  0.162 & 0.253 & 0.541 & 0.865 \\
\cutinhead{($\sigma_r = 2$mmag)}
0--5 days &  0.839 & 0.954 & 0.986 & 0.996\\
5--10 days &  0.182 & 0.462 & 0.912 & 0.985\\
10--20 days &  0.049 & 0.092 & 0.402 & 0.801\\
20--30 days &  0.017 & 0.028 & 0.075 & 0.237 \\
\enddata
\tablecomments{Mean values for $P_{detection}$ for various period ranges (column 1) as function of length of night (Fig. \ref{fig_nightlength}). Assumed parameters are given in the caption of Fig. \ref{fig_nightlength}. For discussion, see \S \ref{night_length}.}
\end{deluxetable}


\subsection{Observing Cadence}
\label{cadence}

Observing cadence is primarily dependent on telescope and detector characteristics  as well as target brightness, with the goals that $\sigma_w$ is minimized, the target remains in the linearity regime of the detector, and the exposure time is not so long as to smear out phase information on any detectable planetary transit. 

\begin{figure*}
\begin{center}
 \includegraphics[angle=270,scale=0.6]{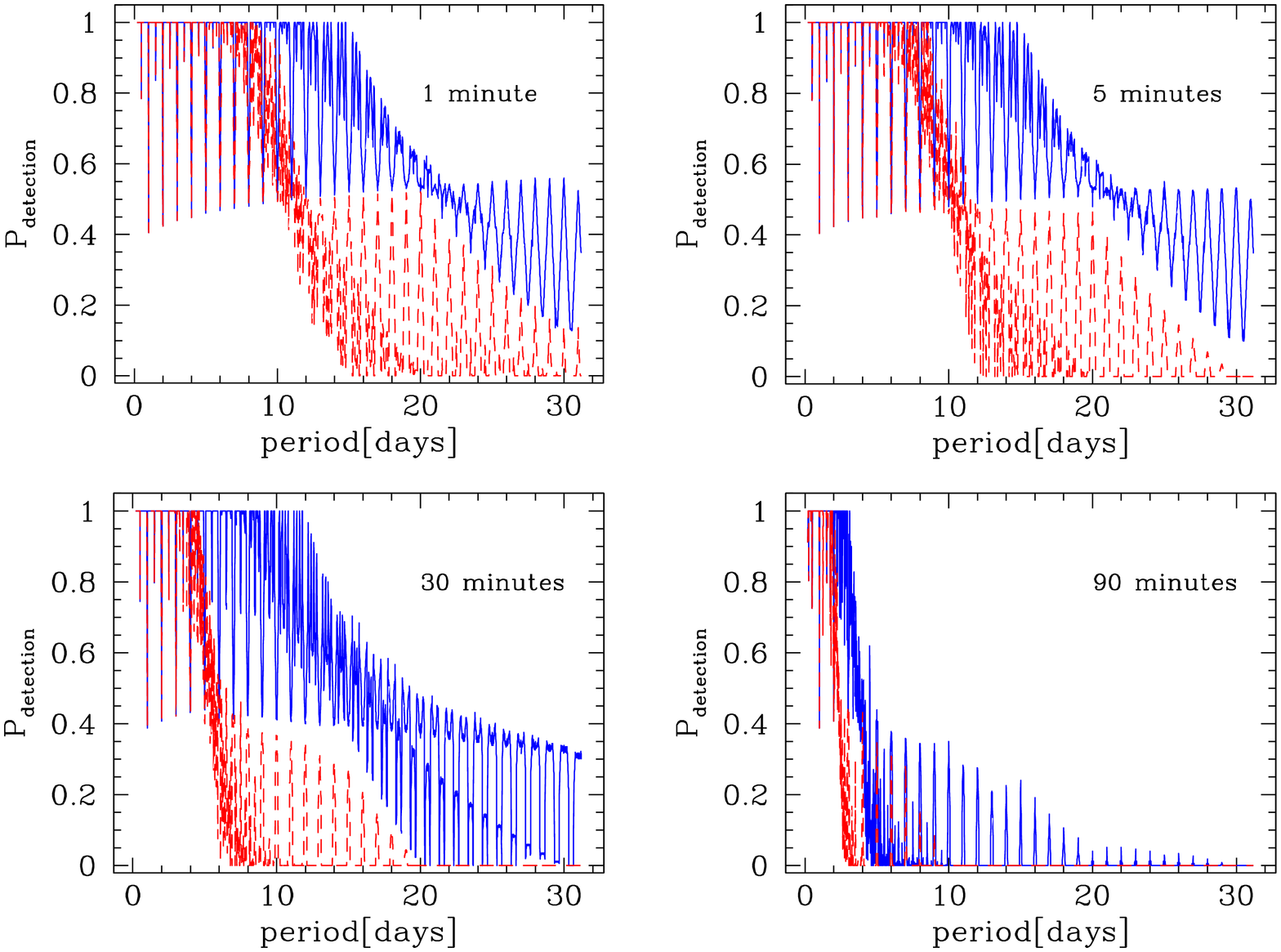} 
 \caption{The influence of observing cadence on the detection efficiency,
 shown for 1, 5, 60, and 90 minutes (indicated in the respective panel).
 $\sigma_w$ is assumed to be 5 mmag. The (blue) solid line indicates the
 detection efficiency $\sigma_r = 0$, and the (red) dashed for $\sigma_r = 2$
 mmag. Additional parameters are: SNR$_{threshold}$ = 7.0, 8 hours of
 observing every night, 60 consecutive nights, $R_{star} = R_{\odot}$,
 $R_{planet} = 0.1 R_{\odot}$, $M_{star} = M_{\odot}$, $M_{planet} \ll
 M_{\odot}$. See Table \ref{tab_cadence} for mean values of $P_{detection}$ over various period ranges, and \S \ref{cadence} for details.}  \label{fig_cadence}
\end{center}
\end{figure*}

Similar to \S \ref{run_length} and \S \ref{night_length}, however, the choice of cadence can also be used as a observing strategy parameter to increase the number of monitored stars at the expense of a lower sampling rate per field (by moving back and forth between fields between exposures, for instance). This effect is simulated in Fig. \ref{fig_cadence}, and values for $P_{detection}$ for different period ranges are given in Table \ref{tab_cadence}, which shows that the effect of changing from a cadence of one to several minutes does not greatly affect the calculated detection probability, especially for very small values of $\sigma_r$. It may therefore be worth considering changing between fields every one or few exposures to increase target number. The effects of red noise produced for such an observing strategy, however, such as flatfielding errors due to the fact that the stars may not be located in exactly the same position in the field as before, are dependent on aspects such as the pointing stability of the telescope used and would need to be explored for the respective observing setup.

\begin{deluxetable}{lcccc}
\tablecaption{Mean $P_{detection}$ Values for Different Observing Cadences \label{tab_cadence}}
\tablewidth{0pc}
\tablehead{
\colhead{Minutes} &
\colhead{1} &
\colhead{5} &
\colhead{30} &
\colhead{90} }
\startdata
\cutinhead{($\sigma_r = 0$)}
0--5 days &  0.984 & 0.983 & 0.974 & 0.725\\
5--10 days &  0.947 & 0.945 & 0.883 & 0.057\\
10--20 days &  0.758 & 0.753 & 0.535 & 0.021\\
20--30 days &  0.431 & 0.425 & 0.264 & 0.001 \\
\cutinhead{($\sigma_r = 2$mmag)}
0--5 days &  0.980 & 0.978 & 0.936 & 0.457\\
5--10 days &  0.864 & 0.819 & 0.214 & 0.015\\
10--20 days &  0.293 & 0.235 & 0.037 & 0.000\\
20--30 days &  0.071 & 0.051 & 0.000 & 0.000 \\
\enddata
\tablecomments{Mean values for $P_{detection}$ for various period ranges (column 1) as function of observing cadence (Fig. \ref{fig_cadence}). Assumed parameters are given in the caption of Fig. \ref{fig_cadence}. For discussion, see \S \ref{cadence}.}
\end{deluxetable}


\subsection{Minimum Number of Sampled Transits}
\label{min_transit_number}


\subsubsection{Two or more Transits}
\label{two_plus}


Transit duration is dependent on period \citep[equation 3 in][]{sm03}. 
Observational cadence determines the number of expected data points in a
single transit. The combination of number of nights in the observing run and
length of a given night sets the expected number of transits sampled during the monitoring campaign.

As explained in \S \ref{transit_number}, we require a minimum of two transits sampled to constitute a detection. Note, however, that other predictions of survey yields in the literature use different numbers for different reasons, e.g., to be able to constrain period, which, with only two transits detected, would be subject to significant aliasing uncertainties, depending on the time elapsed between the two sampled transit events \citep[see \S 7.4 in][]{bls05}. 

\begin{figure*}
\begin{center}
 \includegraphics[angle=270,scale=0.6]{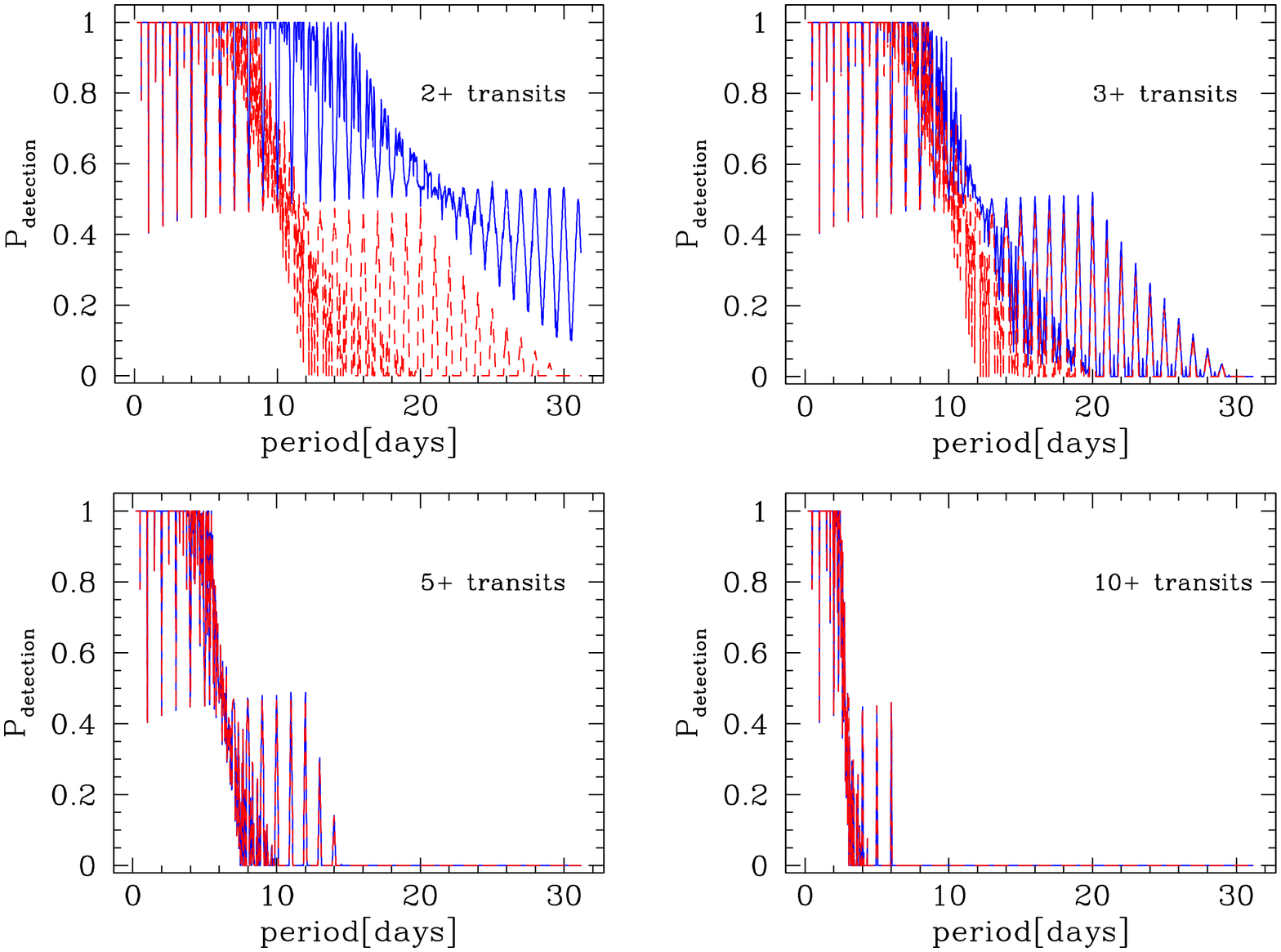} 
 \caption{ The influence of the minimum number of transits sampled on the detection efficiency. Shown is the requirement that 
 (at least) 2, 3, 5, and 10 transits (indicated in the respective panel) be sampled.  $\sigma_w$ is
 assumed to be 5 mmag. The (blue) solid line indicates the detection
 efficiency $\sigma_r = 0$, and the (red) dashed for $\sigma_r = 2$
 mmag. Additional parameters are: SNR$_{threshold}$ = 7.0, a 5-minute
 observing cadence, 8 hours observing per night, 60 consecutive nights,
 $R_{star} = R_{\odot}$,
 $R_{planet} = 0.1 R_{\odot}$, $M_{star} = M_{\odot}$, $M_{planet} \ll
 M_{\odot}$. See Table \ref{tab_sampled_transits} for mean values of $P_{detection}$ over various period ranges, and \S \ref{two_plus} for details.}  \label{fig_sampled_transits}
\end{center}
\end{figure*}

Fig. \ref{fig_sampled_transits} and Table \ref{tab_sampled_transits} show how the detection probability varies as a function of different minimum number of sampled transits. We note that, in this work, only one observation taken during a transit is enough to count this transit as {\it sampled}, but the {\it detection} is still a function of the transit SNR, as explained in \S \ref{algorithm}, as well as the number of sampled transits.

It is interesting to observe that the detection probability of $\sigma_r = 2$ mmag approaches the $\sigma_r = 0$ case for higher minimum number of sampled transits, showing how the "white noise only" case thus becomes increasingly equivalent to the realistic case with red noise present (Fig. \ref{fig_sampled_transits} and Table \ref{tab_sampled_transits} produce identical results for $\sigma_r=0$ and $\sigma_r=2$ mmag for more than five or more transits sampled). In the absence of knowledge of $\sigma_r$, requiring at least three or four detected transits in the data could therefore serve as a alternative for calculating a conservative estimate of survey yield. 

\begin{deluxetable}{lcccc}
\tablecaption{Mean $P_{detection}$ Values for Different Minimum Number of Sampled Transits \label{tab_sampled_transits}}
\tablewidth{0pc}
\tablehead{
\colhead{Min. Number} &
\colhead{2} &
\colhead{3} &
\colhead{5} &
\colhead{10} }
\startdata
\cutinhead{($\sigma_r = 0$)}
0--5 days &  0.983 & 0.981 & 0.965 & 0.549\\
5--10 days &  0.945 & 0.878 & 0.313 & 0.008\\
10--20 days &  0.753 & 0.339 & 0.023 & 0.000\\
20--30 days &  0.425 & 0.066 & 0.000 & 0.000 \\
\cutinhead{($\sigma_r = 2$mmag)}
0--5 days &  0.978 & 0.978 & 0.965 & 0.549\\
5--10 days &  0.819 & 0.819 & 0.313 & 0.008\\
10--20 days &  0.235 & 0.235 & 0.023 & 0.000\\
20--30 days &  0.051 & 0.051 & 0.000 & 0.000 \\
\enddata
\tablecomments{Mean values for $P_{detection}$ for various period ranges (column 1) as function of observing cadence (Fig. \ref{fig_sampled_transits}). Assumed parameters are given in the caption of Fig. \ref{fig_sampled_transits}. For discussion, see \S \ref{two_plus}.}
\end{deluxetable}



\subsubsection{Detections Based on Single Transits}
\label{single_transit}

\begin{figure*}
\begin{center}
 \includegraphics[angle=270,scale=0.6]{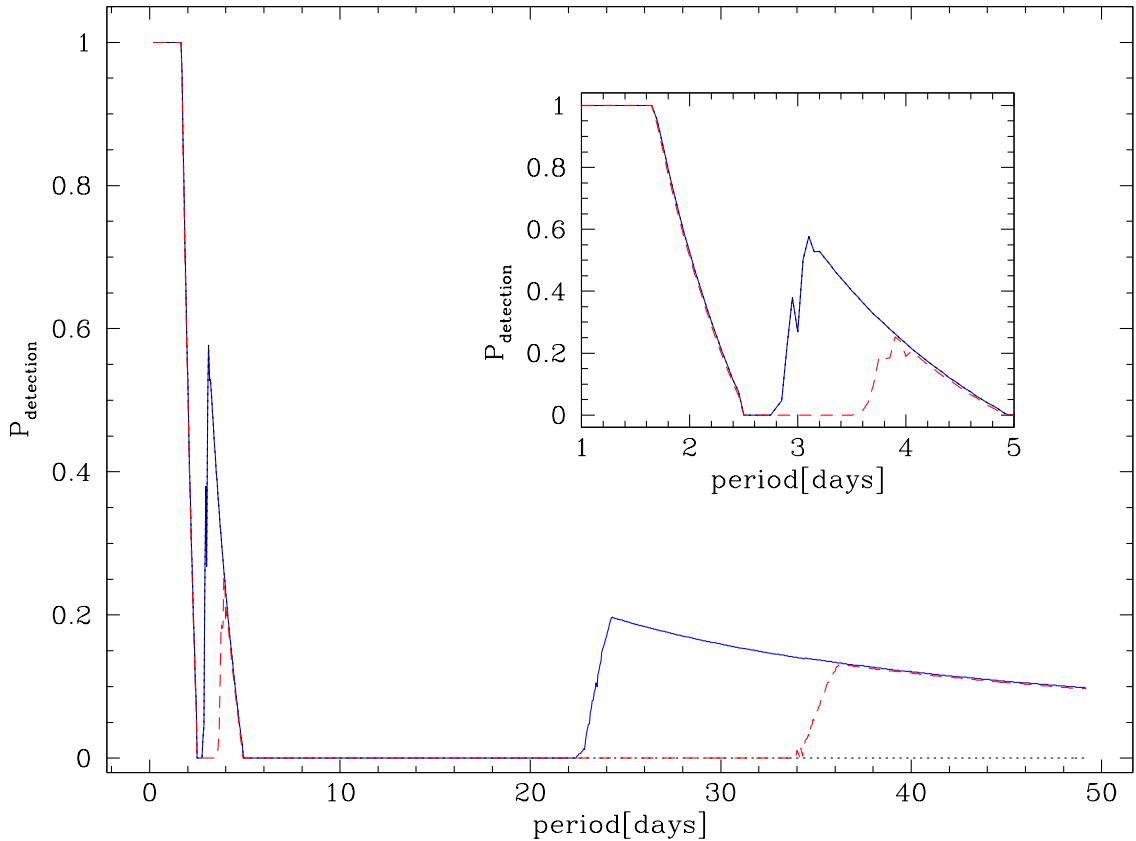} 
 \caption{The behavior of the detection probability as a function of period for a short, white-noise dominated monitoring campaign. $\sigma_w$ is assumed to be 10 mmag. $\sigma_r = 0$ for the solid (blue) and dotted (black) lines; $\sigma_r = 0.5$ mmag for the dashed (red) line. The solid (blue) and dashed (red) lines no longer assume a minimum number of sampled transits, whereas the dotted (black) line requires a minimum of two transits to be present in the data. Additional parameters are: SNR$_{threshold}$ = 7.0, a 7-minute observing cadence, continuous observing (i.e., 24 hours observing per ``night"), 5 consecutive nights, $R_{star} = R_{\odot}$,  $R_{planet} = 0.1 R_{\odot}$, $M_{star} = M_{\odot}$, $M_{planet} = M_{Jupiter}$.  The inset is a zoomed display of the behavior of $P_{detection}$ for periods between 1 and 5 days. See \S \ref{single_transit} for discussion.}  \label{fig_5nights}
\end{center}
\end{figure*}

We now examine the case where it is deemed possible to detect a transit based on a minimum of {\it one} sampled transit. For this scenario, the detection probability may exhibit a non-intuitive
behavior for very short and very long periods. To illustrate these points, we assume 5 ``nights" of uninterrupted 24-hour (e.g., space-based or polar ground-based) observing (see Fig.\ref{fig_5nights}).

Short periods imply short transit durations. For a given cadence, there are few points per sampled transit. At the same time, short periods imply many transits sampled for a given observing run length. With increasing orbital period, the number of sampled transits decreases as the number of data points per transit increases, though not at the same rate. 
This effect can be seen for the short period range in Fig. \ref{fig_5nights}, and it is most intuitively understood by considering the case for $\sigma_r = 0$. Since $\sigma_w = $ transit depth = 10 mmag (Fig. \ref{fig_5nights}), a detection simply requires $SNR_{threshold}^2 = 49$ data points observed during transit (Eq. \ref{simple_equation}), even if they are all located in a single transit. $P_{detection}$ goes to zero at a period of 2.5 days, which, for the stellar and planetary masses and radii given in the caption of Fig. \ref{fig_5nights}, implies a transit length of around 161 minutes. For a 7-minute observing cadence, 23 data points can thus be collected during a single transit. For a 2.5 day period, one would expect to have two transits present in a 5-day observing run (24 hours of observing per ``night"), but two transits would only contain 46 data points, and not the 49 required for $SNR_{transit} > SNR_{threshold}$. $P_{detection}$ in Fig. \ref{fig_5nights} therefore goes to zero at that point. 

For periods longer than 3 days, however, the expected number of data points per transit is 24.5, and thus, sampling two transits somewhere in one's data is sufficient for a detection, resulting in an increase of $P_{detection}$ with period as it approaches 3 days.

An inverse effect can be observed at periods longer than the duration of the observing run and is again most easily explained by considering the $\sigma_r = 0$ case. As expected, $P_{detection}$ drops to zero for periods longer than the observing run, since, although it is possible to sample one transit during the observing run, the length of this one transit is too short for enough data points (49) to be sampled during its duration (and it is obviously impossible to sample more than one transit, as indicated by the dotted line). This situation, however, changes as the period approaches 24 days (transit duration = 343 minutes). From that point on, the transit duration will be long enough to fit 49 data points at a 7-minute cadence. 
Therefore, if a single transit is observed, it is long enough to gather
enough data to fulfill the threshold SNR criterion.  
Ultimately, the probability that any
transit occurs at all during the observing run, and thus the detection
efficiency, approaches zero.




\subsection{Transit Depth and Duration}
\label{transit_duration}

Throughout the paper, we calculate the transit depth and duration according to equations in \citet{sm03}, thereby assuming a central transit, i.e., $i=90^{\circ}$ and $b=0$, as well as solar and Jupiter values for stellar and planetary radii and masses. In this subsection, we set those parameters to certain values that may not be consistent with physical laws, but are meant to illustrate the behavior of the detection probability as a function of transit duration and depth. 

\begin{figure*}
\begin{center}
 \includegraphics[angle=270,scale=0.6]{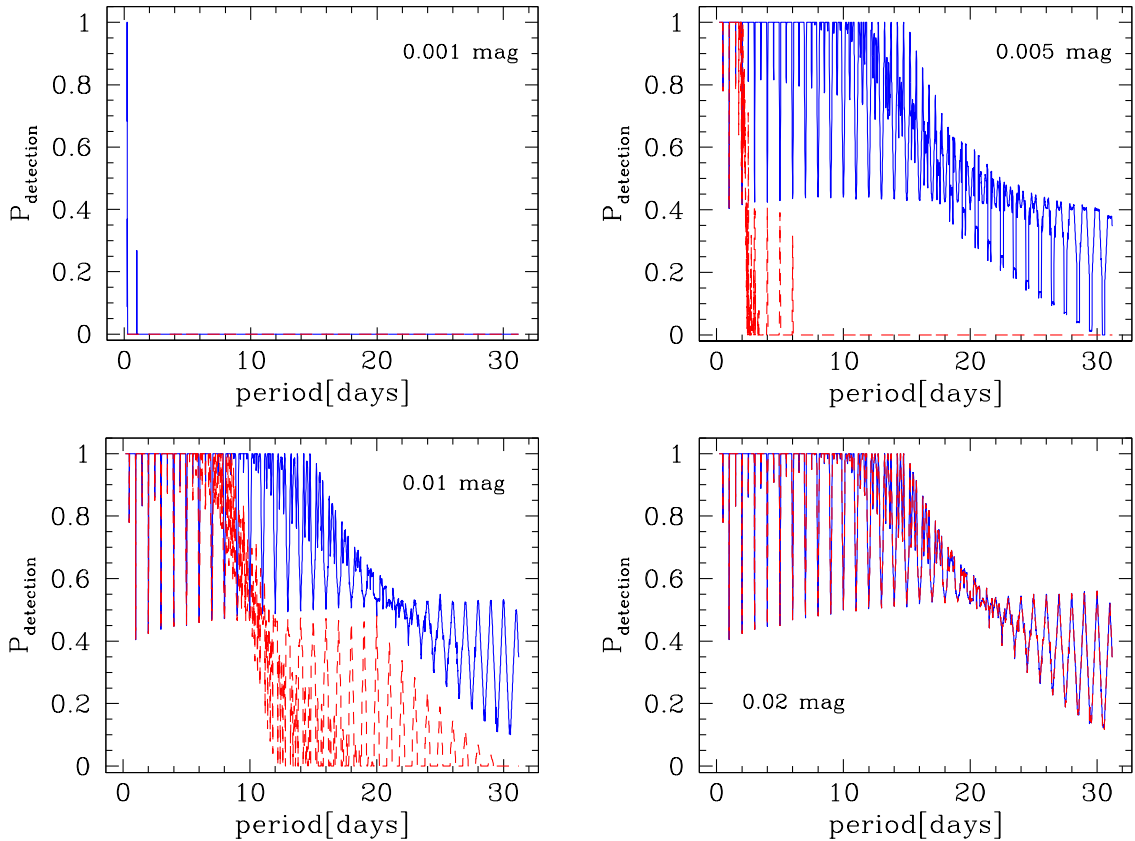} 
 \caption{The influence of the depth of transit on the detection efficiency,
 shown for 1, 5, 10, and 20 mmag (indicated in the respective panel).
 $\sigma_w$ is assumed to be 5 mmag. The (blue) solid line indicates the
 detection efficiency $\sigma_r = 0$, and the (red) dashed for $\sigma_r = 2$
 mmag. Additional parameters are: SNR$_{threshold}$ = 7.0, a 5-minute
 observing cadence, 8 hours observing per night, 60 consecutive nights,
 transit duration calculated using solar radius and mass for the star and
 Jupiter values for planetary mass and radius. See Table \ref{tab_transitdepth} for mean values of $P_{detection}$ over various period ranges, and \S \ref{transit_duration} for details.}  \label{fig_transitdepth}
\end{center}
\end{figure*}

Transit depth is primarily a function of stellar and planetary radius, which we set to solar and Jupiter values before, resulting in a transit depth of 0.01 mag or 1\% of the relative flux. In order to show how much more challenging the detection of smaller planets is, or conversely, how much easier the detection of larger planets is for given observing parameters, we vary transit depth in Fig. \ref{fig_transitdepth} and show mean $P_{detection}$ values in Table \ref{tab_transitdepth}. 

For a parent star with $R=R_{\odot}$, the panels represent planets of 0.3 $R_{Jupiter}$ (top left), 0.7 $R_{Jupiter}$ (top right), 1.0 $R_{Jupiter}$ (bottom left), and 1.4 $R_{Jupiter}$ (bottom right). Note that an Earth-sized planet would have a radius of around 0.1 $R_{Jupiter}$ and produce an eclipse around a solar-sized star with depth of 10\% of that assumed in the top left panel, i.e., 0.0001 mag. It is worth pointing out how significant the difference in detection probability is for shallow transits between $\sigma_r = 0$ and $\sigma_r \neq 0$, substantiating the claim that space-based observing is necessary to find very small planets (see also \S \ref{space}), as recently evidenced by the discovery of CoRoT-7b (L\'{e}ger et al. in preparation; Rouan et al. in preparation; Bouchy et al. in preparation). 

\begin{figure*}
\begin{center}
 \includegraphics[angle=270,scale=0.6]{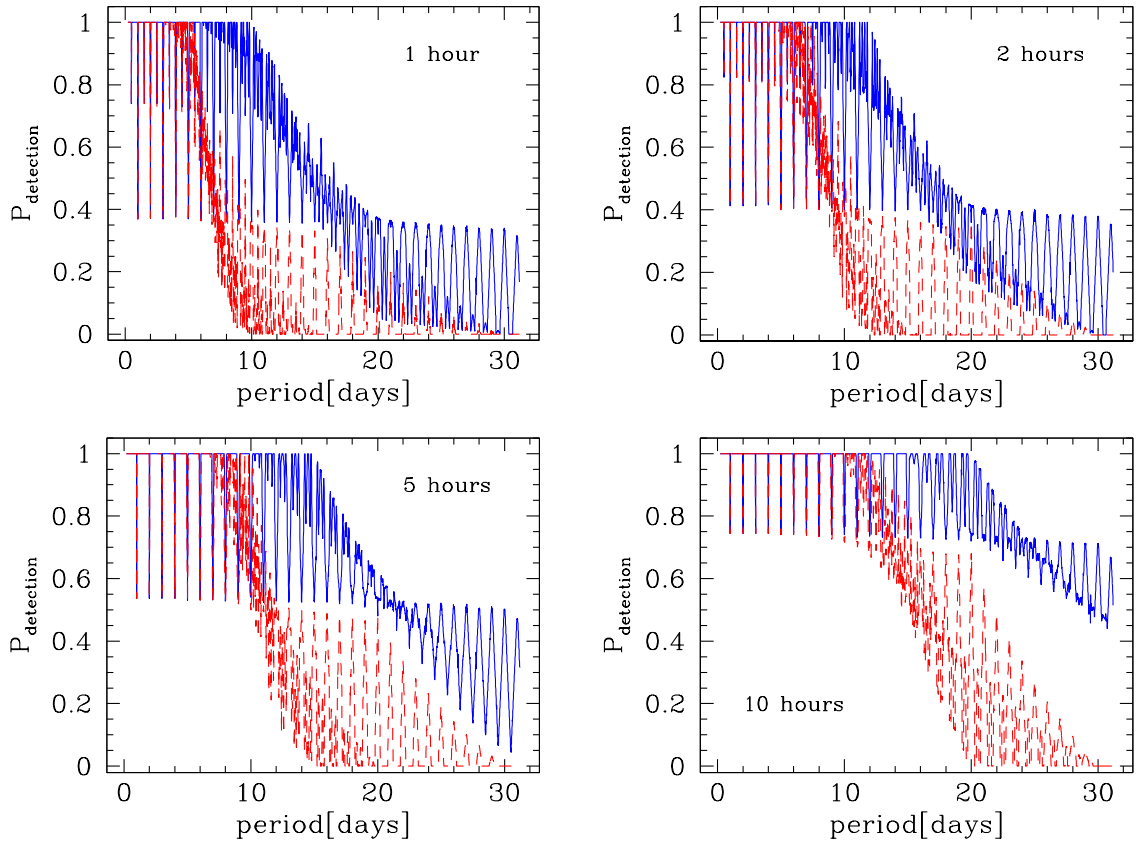} 
 \caption{The influence of transit duration on the detection efficiency, shown
 for 1, 2, 5, and 10 hours (indicated in the respective panel).  $\sigma_w$ is
 assumed to be 5 mmag. The (blue) solid line indicates the detection
 efficiency $\sigma_r = 0$, and the (red) dashed for $\sigma_r = 2$
 mmag. Additional parameters are: SNR$_{threshold}$ = 7.0, a 5-minute
 observing cadence, 8 hours observing per night, 60 consecutive nights,
 transit depth = 0.01 mag. See Table \ref{tab_transitduration} for mean values of $P_{detection}$ over various period ranges, and \S \ref{transit_duration} for details.}  \label{fig_transitduration}
\end{center}
\end{figure*}

Transit duration is a function of orbital period, $i$, and stellar and planetary masses and radii. Rather than following the physical dependence on period, we set transit duration to fixed values of 1, 2, 5, and 10 hours in the four panels of Fig. \ref{fig_transitduration} (see also Table \ref{tab_transitduration}), thereby still assuming values for the parameters mentioned in the figure caption, including a transit depth of 0.01 mag. While longer transits are obviously easier to detect, the increase of detectability with transit duration is slow but sensitively dependent on $\sigma_r$. 

\begin{deluxetable}{lcccc}
\tablecaption{Mean $P_{detection}$ Values for Different Transit Depths \label{tab_transitdepth}}
\tablewidth{0pc}
\tablehead{
\colhead{Transit Depth} &
\colhead{1 mmag} &
\colhead{5 mmag} &
\colhead{10 mmag} &
\colhead{20 mmag} }
\startdata
\cutinhead{($\sigma_r = 0$)}
0--5 days &  0.009 & 0.977 & 0.983 & 0.984\\
5--10 days &  0.000 & 0.916 & 0.945 & 0.947\\
10--20 days &  0.000 & 0.677 & 0.753 & 0.755\\
20--30 days &  0.000 & 0.355 & 0.425 & 0.428 \\
\cutinhead{($\sigma_r = 2$mmag)}
0--5 days &  0.000 & 0.435 & 0.978 & 0.983\\
5--10 days &  0.000 & 0.005 & 0.819 & 0.947\\
10--20 days &  0.000 & 0.000 & 0.235 & 0.755\\
20--30 days &  0.000 & 0.000 & 0.051 & 0.428 \\
\enddata
\tablecomments{Mean values for $P_{detection}$ for various period ranges (column 1) as function of transit depth (Fig. \ref{fig_transitdepth}). Assumed parameters are given in the caption of Fig. \ref{fig_transitdepth}. For discussion, see \S \ref{transit_duration}.}
\end{deluxetable}

\begin{deluxetable}{lcccc}
\tablecaption{Mean $P_{detection}$ Values for Different Transit Durations \label{tab_transitduration}}
\tablewidth{0pc}
\tablehead{
\colhead{Transit Duration} &
\colhead{1 hour} &
\colhead{2 hours} &
\colhead{5 hours} &
\colhead{10 hours} }
\startdata
\cutinhead{($\sigma_r = 0$)}
0--5 days &  0.976 & 0.981 & 0.989 & 0.996\\
5--10 days &  0.896 & 0.920 & 0.957 & 0.987\\
10--20 days &  0.506 & 0.585 & 0.776 & 0.938\\
20--30 days &  0.193 & 0.240 & 0.409 & 0.690 \\
\cutinhead{($\sigma_r = 2$mmag)}
0--5 days &  0.964 & 0.975 & 0.986 & 0.995\\
5--10 days &  0.414 & 0.676 & 0.879 & 0.974\\
10--20 days &  0.058 & 0.126 & 0.272 & 0.609\\
20--30 days &  0.010 & 0.028 & 0.050 & 0.121 \\
\enddata
\tablecomments{Mean values for $P_{detection}$ for various period ranges (column 1) as function of transit duration (Fig. \ref{fig_transitduration}). Assumed parameters are given in the caption of Fig. \ref{fig_transitduration}. For discussion, see \S \ref{transit_duration}.}
\end{deluxetable}










\section{Application and Examples}
\label{surveys}

The examples used in \S \ref{parameters} to illustrate the influences of various observing strategy and astrophysical parameters on the transit detection probability resemble observing campaigns typical of the very successful wide-field transit surveys such as HAT, TrES, XO, SWASP, etc \citep[e.g., ][]{bnk07,ocm06,msv06,psc06}. In contrast, this section shows examples and consequences of observational window functions for fundamentally different setups of monitoring projects. 


 



\subsection{Space Based Surveys}
\label{space}

Compared to ground-based counterparts, space-based transit surveys such as CoRoT \citep{bab06} 
have the principal two advantages that (a) they are not subject to interruptions in observing due to the diurnal cycle (see \S \ref{night_length}), and that (b) they do not need to deal with the Earth's atmosphere (see \S \ref{red_white_noise}). The latter aspect in particular makes them the currently only realistic option of detecting Earth-sized planets around sun-like stars, which is one of the explicit goals of the recently launched Kepler Mission \citep{bkb09}. 

\begin{figure*}
\begin{center}
\plotone{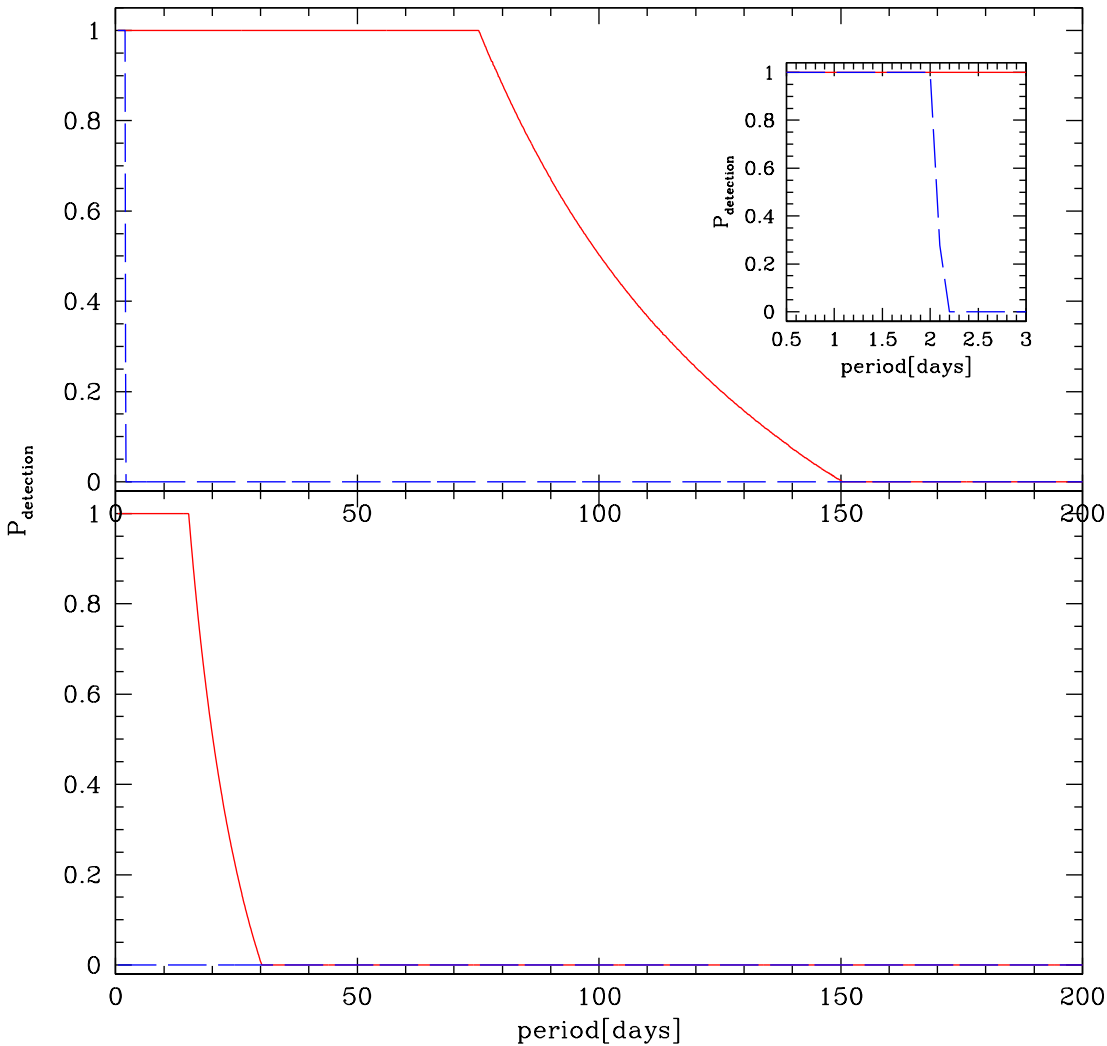} 
 \caption{Simulated space survey loosely based on the CoRoT short runs (bottom panel) and long runs (top panel). The solid (red) line indicates the detection probability of a Jupiter-sized planet around a solar-type star, the dashed (blue) line indicates the detection probability of the recently discovered exoplanet CoRoT-7b around its (K0 dwarf) parent star (transit depth $\sim$ 0.5 mmag; period $\sim$ 0.9 days). The inset in the top panel represents a zoomed display of the 0.5--3.0 day period range. Parameter values assumed in this simulation are: $SNR_{threshold} = 7.0, \sigma_w = 1$ mmag, $\sigma_r=0.5$ mmag, continuous observing with a 15-min cadence over 30 days (bottom panel) and 150 days (top panel). 
  See \S \ref{space} for details.}  \label{fig_corot}
\end{center}
\end{figure*}

Fig. \ref{fig_corot} shows the detection probability for simulated space-based surveys of various lengths, loosely modeled after the long and short observing runs by the CoRoT satellite, thereby assuming somewhat generic parameters for survey strategy and photometric precision (see caption). The solid and dashed lines respectively indicate the detection probabilities for a Jupiter-sized planet and for the recently discovered exoplanet CoRoT-7b around its parent star, a K0 dwarf (transit depth $\sim$ 0.5 mmag; period $\sim$ 0.9 days). We note that our simulations of an Earth-sized planet around a solar-type star produce a detection probability of zero for all periods. 

\subsection{Synoptic Surveys}
\label{synoptic}

Synoptic surveys typically provide high-quality photometric time-series data of very low cadence but over extended periods of time. Thus, they are not primarily designed to find planetary transits but nevertheless present data sets that are worth probing for their existence \citep[see for instance][]{pjk08}. In fact, several transiting planets have been discovered {\it a posteriori} in the Hipparcos archives such as HD 209458b \citep{ra00} and HD 189733b \citep{hl06}.

\begin{figure*}
\begin{center}
\plotone{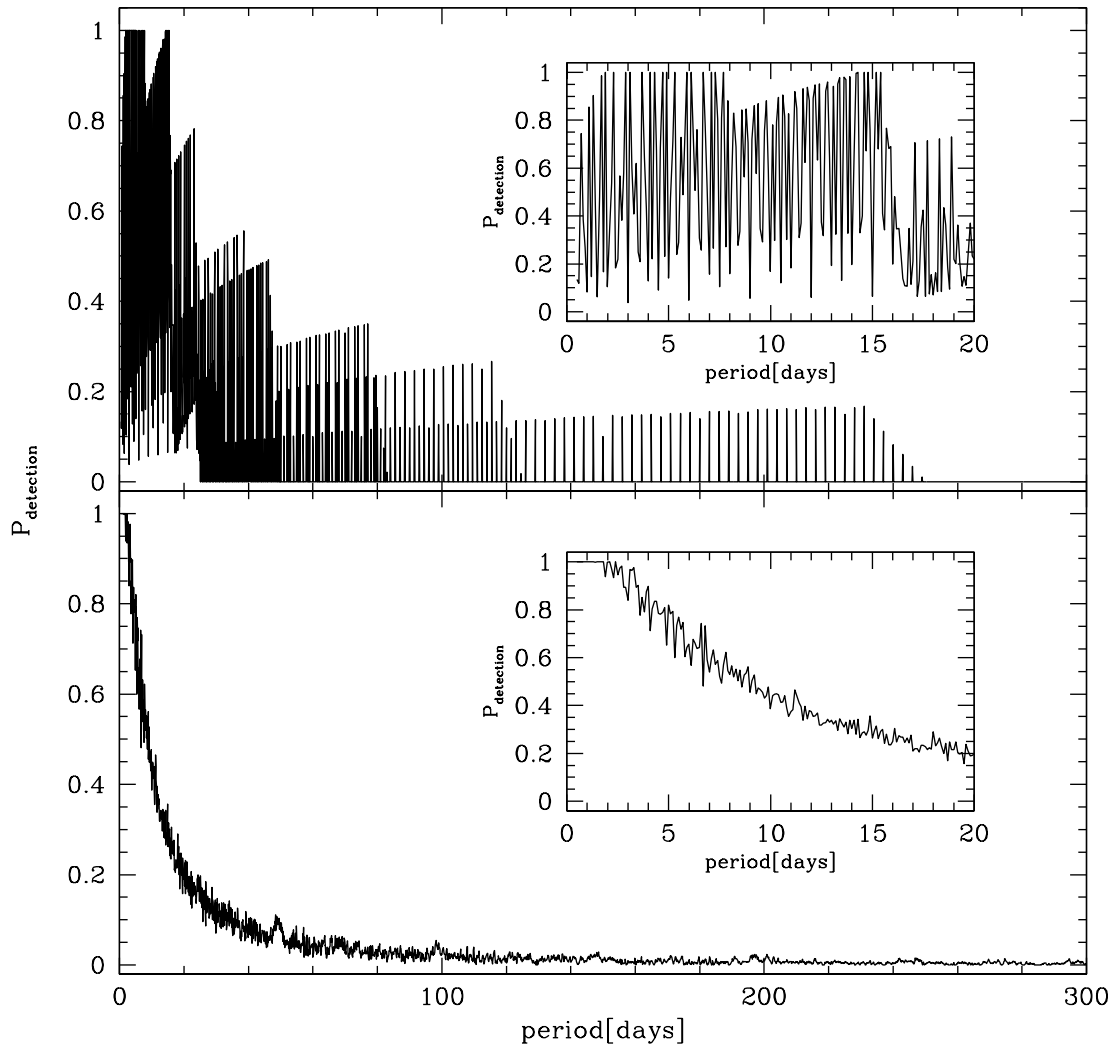} 
 \caption{Simulated synoptic surveys loosely modeled after LSST (top panel) and Hipparcos (bottom panel). The insets are zoomed regions for the shorter periods. We assume solar-type stars and Jovian planets in these simulations. The 1,000 observations in the top panel are obtained over eight years, whereas the 190 observations in the bottom panel are taken over three years. See \S \ref{synoptic} for details.}  \label{fig_synoptic}
\end{center}
\end{figure*}

The panels in Fig. \ref{fig_synoptic} are produced by observational window functions of synoptic surveys loosely based on the (the future, ground-based) Large Synoptic Survey Telescope \citep[LSST;][]{ita08} in the top panel and (the space-based) Hipparcos mission \citep{p97} in the bottom panel. For both panels, we require a $SNR_{threshold}=7.0$, at least two sampled transits, and assume solar and Jupiter values for stellar and planetary mass and radius. The principal differences between the two window functions in Fig. \ref{fig_synoptic} are due to the different assumptions in $\sigma_w$ (5 mmag for top panel; 1.5 mmag for bottom panel) and $\sigma_r$ (1 mmag for top panel; 0.5 mmag for bottom panel), and the different number of data points obtained over different lengths of time. 

For the top panel, we assumed a cadence of a single, 30-second exposure time image every three nights, accumulated over around eight years, such that the total number of images is 1,000. The mean value for $P_{detection}$ over various period ranges are as follows: 
\begin{itemize}
\item 0--10 days: $<P_{detection}>$ = 0.555;
\item 10--50 days: $<P_{detection}>$ = 0.239;
\item 50-100 days: $<P_{detection}>$ = 0.025;
\item 100--200 days: $<P_{detection}>$ = 0.007;
\end{itemize}

For the bottom panel, we chose a cadence based on the actual observations of a Hipparcos star with 190 epochs, downloaded from the NASA Star and Exoplanet Database\footnote{http://nsted.ipac.caltech.edu}. Basically, the 190 observations were obtained over three years in groups of several images every few tens of days. The mean value for $P_{detection}$ over various period ranges are as follows: 
\begin{itemize}
\item 0--10 days: $<P_{detection}>$ = 0.746;
\item 10--50 days: $<P_{detection}>$ = 0.156;
\item 50-100 days: $<P_{detection}>$ = 0.034;
\item 100--200 days: $<P_{detection}>$ = 0.012;
\end{itemize}

Finally, it should be noted that we ran the equivalent simulations to the ones in Fig. \ref{fig_synoptic}, but thereby assuming an Earth-sized planet instead of a Jupiter-sized one. Both detection probabilities were identical to zero for all periods. 

\section{Discussion and Conclusion}
\label{conclusion}




This work quantitatively illustrates the influence of a number of observing strategy and astrophysical parameters on the detection efficiency of existing planetary transits as a function of period under general assumptions listed in \S \ref{description_algorithm} and parameters given in the various figure captions. The influences of red and white noises upon this detection efficiency are first examined in their own right, and then included in every simulation of the aforementioned parameters. Red noise is confirmed to be the dominant challenge to overcome in the search for planetary transits, as seen in the discussion in \S \ref{red_white_noise}, Figures \ref{fig_rednoise} and \ref{fig_whitenoise}, and Tables \ref{tab_rednoise} and \ref{tab_whitenoise}. All parameters being equal, a factor 4 increase in $\sigma_r$ produces a much more significant reduction in $P_{detection}$ than a factor 10 increase in $\sigma_w$.

In particular, we explicitly address controllable strategy parameters such as the number of nights for which one may choose to monitor a given target field, the number of hours per night one may stay on this field, and the observational cadence with which the field is monitored. We furthermore examine the influence of astrophysical parameters on detection efficiency, such as transit depth and duration. Finally, we look at parameters typically involved in the calculation of the projected yield of a given transit survey such as the minimum number of transits required for detections, and illustrate two non-intuitive effects that occur when the criterion of a minimum number of sampled transits is abandoned and detection is based only on SNR. Along with visualization of the effects caused by the various parameters in the figures, we provide quantitative means of comparison for different period ranges in the accompanying tables.

A consideration that did not factor into the calculation of $P_{detection}$ is the fraction of data points outside of transit (we assumed that this fraction is much higher than the number of points sampled in transit). In order to detect a transit
in one's data, one needs to have both brightness levels well measured such that the difference between them becomes signficant enough to enable a detection. For instance, an observing run that {\it only} obtains data during
transit would, by the metrics used in this paper, detect the transit, provided the SNR is high enough. In real life,  however, the data would appear perfectly flat and no sensible algorithm would flag the signal as a possible planetary transit. Obviously, sparse cadences are more susceptible to this admittedly pathologic pitfall than well sampled ones. One scenario where one may encounter a problem like this would be in the attempt to detect transits among long-period planets discovered by radial velocity work, which would tend to exhibit long transit durations. 

More generally, we caution that the detection of an existing transit in ``real life'' is dependent on a large number of properties of the data reduction and analysis pipeline and transit detection methods, including human experience and human error potential, which cannot possibly be parametrized as an ensemble or included in any code.  Therefore, the significance of our results and predictions, although quantitative, are necessarily subject to an unknown fudge or scaling factor. As pointed out by \citet{bg08}, the non-uniformity of the definitions of detection criteria cause the largest uncertainty in transit survey yield predictions. 

Nevertheless, we specifically allowed for parameters that are typically calculable in transit survey designs to be used as input to the code in order to make it as practically applicable as possible. Consequently, even in the presence of the unknown fudge factor mentioned above, comparisons between different observing strategies is quantitatively possible to optimize survey yield. For instance, under some circumstances it appears much more favorable to increase the observational cadence to add a second monitoring field to one's project than switching fields in the middle of the night or in the middle of the observing run (see \S\S \ref{run_length}, \ref{night_length}, and \ref{cadence}), provided it is possible to repeatedly achieve very good pointing of the telescope to reduce the additional red noise component that might arise from flatfielding errors otherwise. Thus, it may well be advisable for transit surveys to consider this trade-off between cadence and fields monitored as it can lead to a dramatic change in the predicted planet yield of the survey.

Furthermore, the examination of the effects of red and white noise in \S \ref{red_white_noise} and throughout the paper give quantitative insight into what size of planet one may expect to realistically detect in one's data, given observing strategy parameters. In general, the depth of transit one may hope to detect in one's data needs to be larger than the magnitude of $\sigma_r$, as seen in \S \ref{transit_duration} and evidenced Figures \ref{fig_transitdepth} and \ref{fig_corot}. This is confirmed very well in Fig. \ref{fig_corot}, showing that the detection of CoRoT-7b around its parent star, given their sizes and orbital period, are right at the limits of the CoRoT satellite for a single long run (i.e., without combining data from several runs). 




The code used for all calculations in this paper is available from KvB upon request.


\acknowledgements
We thank S. Seager, G. Mall\'en-Ornelas, and B. L. Lee for many helpful discussions about
window functions, and F. Pont for invaluable assistance with red noise
considerations. We furthermore express our thanks to R. Alonso, J. Pepper, and B. S. Gaudi for sharing insights into their ground-based and space-based data with respect to red noise characteristics and decorrelation timescales. Finally, we extend our gratitude to the anonymous referee for comments, encouragement, and a very insightful suggestion that noticeably improved the quality of the manuscript, as well as the scientific editor for pointing out a number of shortcomings with respect to mentioning and giving credit to the much more rigorous treatment of red noise in the mathematics and statistics literature.





\bibliography{wf}


\end{document}